\documentclass[dvips]{article}
\usepackage{icrctc07}

\title{Observations of the Crab Nebula with the Whipple 10 m Telescope}
\shorttitle{Crab Nebula with the Whipple 10 m}
\authors{J. Grube$^{1}$ for the VERITAS Collaboration.}
\shortauthors{J. Grube et al.}
\afiliations{$^1$School of Physics and Astronomy, University of Leeds, Leeds,
  West Yorkshire, LS2 9JT, United Kingdom\\
$^2$For full author list see G. Maier, ``VERITAS: Status and Latest Results'',
  these proceedings}
\email{jg@ast.leeds.ac.uk}
\abstract{
Due to the strong and steady TeV 
$\gamma$-ray emission from the Crab Nebula supernova remnant, its measured
flux and energy spectrum can be used to verify the calibration and data
reduction methods applied to IACT data acquired over many observing
seasons. This gives us confidence in the results obtained on variable TeV sources
observed over the same period and in relating the sensitivity of new
instruments to historical datasets. Here we present the results of an analysis
of 65.3 hours of good quality data taken on the Crab Nebula between October
2000 and March 2006 with the Whipple 10m telescope. The total exposure resulted in a 46 $\sigma$ signal with 11886 selected excess events. 
The energy spectrum was best fit by a power law of the form dN/dE $=$ 
(3.19 $\pm$ 0.07$_{\rm{stat.}}$)  $\times10^{-11} \cdot \left(\rm{E}/{1 \rm{TeV}}\right)^{-2.64 
\pm 0.03_{\rm{stat.}}} \rm{cm}^{-2} \rm{s}^{-1} \rm{TeV}^{-1}$ in the energy range
0.49--8 TeV. The systematic uncertainty in the flux 
was estimated to be 30\%, with a systematic error of 0.2 in the photon index. A reasonable agreement is 
shown for a fit to a constant flux over the 6 years.
}

\begin{document}
\maketitle
%Begin the section.
%%%%%
\section{Introduction}
%%%%
The Crab Nebula supernova remnant has served as the TeV $\gamma$-ray ``standard candle'' for
Imaging Atmospheric Cherenkov Telescopes (IACTs) since its successful
detection with a 37 pixel camera on the Whipple 10m telescope in 1989 $[1,
  2, 3, 4]$. Here we provide a record of its flux and energy spectrum from 2000
to 2006 as observed with the Whipple 10m telescope in its current configuration.

The aim of this work is to prove the suitability of various data
reduction and analysis techniques and the stability of the instrument for
long-term source monitoring. The full study will appear in $[5]$. Here we
compare the results obtained with two
sets of Hillas image parameter-based background rejection cuts, ``Hard cuts''
and ``Loose cuts''. We discuss errors in energy reconstruction and the variation in effective
collection area with the zenith angle of observations, both of which were quantified using
simulations. Finally, we plot three integral flux measurements per observing
season as a Crab Nebula light curve, and a differential energy spectrum obtained from
the combined data set is presented alongside results from HEGRA, H.E.S.S. and
MAGIC. 

\section{Observations}
%%%%%

%%%%%
\begin{table}[h]
\begin{center}
\begin{tabular}{ccccc}
\hline
\hline
Obs. Period & T$_{\rm{L}}$ & N$_{\rm{Runs}}$ & $\langle\Theta\rangle$ & $\langle\rm{Rate}\rangle$ \\ 
  (Month/Year)   & (hr) &   & (deg.) & (Hz) \\
\hline                    
 10/00 - 01/01 & 13.2 & 29  & 18.8 & 29.7 \\
 10/01 - 03/02 & 23.9 & 52  & 17.1 & 28.0 \\
 11/02 - 02/03 &  8.3 & 18  & 17.2 & 17.6 \\
 11/03 - 03/04 &  5.5 & 12  & 17.8 & 23.5 \\
 10/04 - 04/05 &  5.5 & 12  & 15.0 & 22.0 \\ 
 10/05 - 03/06 &  8.8 & 19  & 18.6 & 22.8 \\
     Total       & 65.3 & 142 & 17.6 & 25.5 \\
\hline
\end{tabular}
\caption[Whipple 10 m telescope observations of the Crab Nebula]
{Observations of the Crab Nebula. T$_{\rm{L}}$ is the total time ON
  source,$\langle\rm{Rate}\rangle$ is the mean raw trigger rate.
}\label{TabCrabOb}
\end{center}
\end{table}
%%%%%%%%%

The Whipple 10 m telescope 
%% is situated on Mt.~Hopkins, Arizona at an elevation of 2312 m and 
is operating with a 379 pixel, $2.6^{\circ}$ field of
view camera. This is the high-resolution inner section of the 490 pixel camera installed in
1999 $[6]$; the 111 outer guard ring pixels were decommissioned in 2003 and
consequently have not been included in this analysis.

The Crab Nebula datasets used are listed by season in Table \ref{TabCrabOb}. For
this systematic study only 28 minute observation runs centered 
on the source position (ON runs) followed by a matching OFF source run were
included. The data cover a range in zenith angle $\Theta$ of 
10--30$^{\circ}$. Only data taken under good weather conditions, for which the RMS spread in the raw telescope
trigger rate was less than 1.5 Hz were included. As can be seen in Table
\ref{TabCrabOb}, the average $raw$ trigger rate varies somewhat between
observing seasons, which is indicative of small changes in telescope efficiency probably due to a combination of instrumental and environmental factors.

\section{Data Reduction}

\begin{table*}
\begin{center}
\begin{tabular}{lcccccc|cc}
\hline
\hline
Cut & Size & Distance & RSL & RSW & Alpha  & Len./Size & $\sigma$/$\sqrt{\rm{hr}}$ & R$_{\gamma}$ \\
    & (pe) & (deg.)   &     &     & (deg.) & (deg./pe) &   & (min.$^{-1}$) \\
\hline
\hline
Hard   & $>$80  & 0.2 - 0.95 & -2.0 - 1.6 & -2.0 - 1.6 & $<$15 & $<$0.0011 & 5.26 & 3.12 $\pm$ 0.18 \\
Loose  & $>$80  & 0.2 - 0.95 & -2.0 - 2.0 & -2.0 - 2.0 & $<$22 &  --       & 3.56 & 5.51 $\pm$ 0.49 \\
\hline
\hline
\end{tabular}
\caption[Whipple 10 m selection cuts]
{Image selection cuts. The significance $\sigma$/$\sqrt{\rm{hr}}$ and rate 
R$_{\gamma}$ of selected events are from 10 hours of observations of the Crab Nebula during 2000--2006 
at $\sim$20$^{\circ}$ zenith angle.
}\label{TabCutsQ}
\end{center}
\end{table*}

\begin{figure}
\begin{center}
\includegraphics [width=0.4\textwidth]{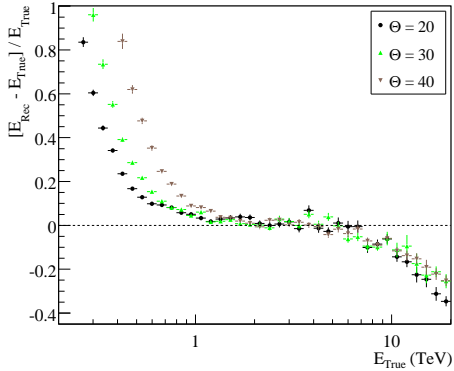}
\vspace{-0.7cm}
\end{center}
\caption{Mean relative error in reconstructed \mbox{$\gamma$-ray} energy as a function of simulated 
energy E$_{\rm{True}}$ at 20$^{\circ}$, 30$^{\circ}$ ,and 40$^{\circ}$ zenith angle.}\label{fig1}
\end{figure}

We applied the Islands method of $[7]$ to extract clean Cherenkov images. The
images were parameterized according to image intensity, shape and
orientation. The Length and Width were converted to the ``reduced scaled''
parameters RSL and RSW $[2]$.

A subset of 10 hours of the Crab Nebula observations from 2000--2006 recorded
at $\sim$20$^{\circ}$ zenith angle were used to 
select two sets of $\gamma$-ray selection cuts: Loose cuts and Hard cuts, the latter for measuring a $\gamma$-ray flux and energy spectrum 
with a high detection significance (following equation 17 of $[8]$). The number of excess $\gamma$-ray type events 
\mbox{N$_{\gamma}$ $=$ N$_{\rm{ON}}$ $-$ $\alpha$ $\cdot$ N$_{\rm{OFF}}$} is
calculated from number of events passing the cuts, scaled by the ratio of the
ON to OFF source exposure time $\alpha$. Table \ref{TabCutsQ}
lists the multi-parameter selection cuts, corresponding significance and
$\gamma$-ray detection rate (defined as \mbox{R$_{\gamma}$ $=$ N$_{\gamma}$ /
T$_{\rm{ON}}$}). 
%%%%%%

\begin{figure}
\begin{center}
\includegraphics [width=0.45\textwidth]{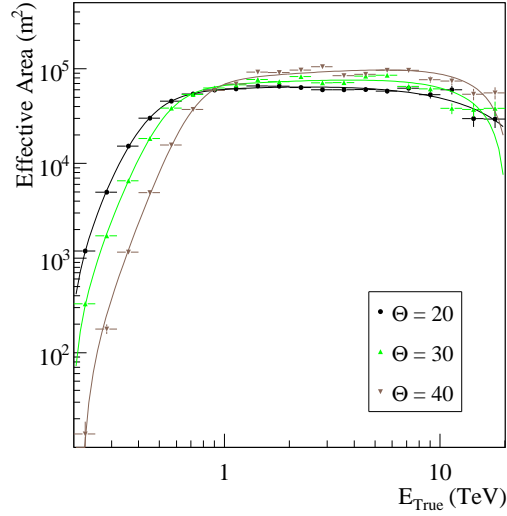}
\vspace{-0.7cm}
\end{center}
%%\vspace{-0.8cm}
\caption{Effective collection area versus true simulated energy after Hard cuts.}\label{fig2} 
\end{figure}

\section{Energy Evaluation and Effective Area}

Simulated $\gamma$-rays at four zenith angles and three telescope efficiencies $\mu$ were used to fill lookup tables for the mean simulated energy E$_{\rm{True}}$, Length, 
and Width as a function of Distance and Log(Size).
A two-dimensional Gaussian smoothing function was applied to 
the lookup tables with $\sigma_{\rm{Dist}}$ $=$ 0.05$^{\circ}$ in Distance and $\sigma_{\rm{Log(S)}}$ $=$ 
0.01 in Log(Size).
For each event, the reconstructed energy E$_{\rm{Rec}}$ was calculated from the lookup tables by linearly 
interpolating between $\mu$ and $\cos\Theta$. Figure \ref{fig1} shows the mean relative error of the reconstructed energy (E$_{\rm{Rec}}$ $-$ 
E$_{\rm{True}}$) / E$_{\rm{True}}$ as a function of simulated energy E$_{\rm{True}}$ at zenith angles of 
20$^{\circ}$, 30$^{\circ}$, and 40$^{\circ}$ after applying Hard selection
cuts. At low energies, 
E$_{\rm{Rec}}$ is overestimated due to events with intensity near the telescope trigger threshold. 
A usable energy range above E$_{\rm{Safe}}$ with a relative error of $<$ 10\% was determined for each zenith 
angle, with the minimum being E$_{\rm{Safe}}$ $=$ 0.5 TeV at zenith angle
20$^{\circ}$.

%%%%%%
In order to account for biases in energy reconstruction, the energy spectrum is measured using the effective 
area as a function of reconstructed energy. The 
maximum distance R$_{\circ}$ $=$ 400 m of the simulated air showers from the telescope was chosen to 
encompass the full impact parameter range of triggered events (0--270 m).
The total number 
of simulated $\gamma$-rays is represented by N$_{\rm{sim.}}$(E,$\Theta$,$\mu$), and the number of 
detected events passing selection cuts as N$_{\rm{sel.}}$(E,$\Theta$,$\mu$). The effective areas were fitted 
with an analytical function modified from equation 3 of $[9]$.

Figure \ref{fig2} shows the effective areas
after Hard cuts as a function of
true simulated energy E$_{\rm{True}}$ at
zenith angles $\Theta$ of 20$^{\circ}$, 
30$^{\circ}$, and 40$^{\circ}$ with telescope efficiency 85\%.

\begin{figure}
\begin{center}
\includegraphics [width=0.4\textwidth]{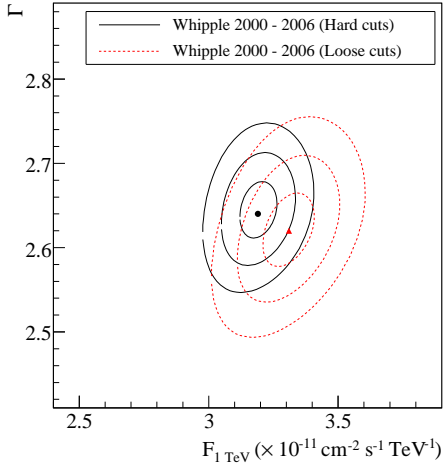}
\vspace{-0.7cm}
\end{center}
\caption{Contour plot of the 68\%, 95\% and 99.9\% confidence intervals from
  the $\chi^2$ fit to a power law for the total 2000-2006 dataset after applying Hard
  cuts or Loose cuts.}\label{fig3}
\end{figure}

The integral $\gamma$-ray flux above a chosen energy threshold is: 
 \[ \rm{F}_{>\rm{E}_{th}} = - \frac{(\frac{dF}{dE})_{th}}{(1-\Gamma)} \cdot 
\rm{E}_{th}^{(2-\Gamma)} \] where a fixed source spectrum with power law 
photon index $\Gamma$ is assumed. 
The excess number of events N$_{\gamma}$ is given by:\\
{\small  $\rm{N}_{\gamma}$ = 
{ ${\rm\frac{dF}{dE}}$ $\int^{\rm{E}_{\rm{max}}}_{\rm{E}_{\rm{min}}}\int^{\rm{T}_{\rm{L}}}_{0} 
A_{\rm{eff}}(\rm{E},\Theta(\rm{t}),\mu(\rm{t})) \cdot \rm{E}^{-\Gamma}\,
\rm{dt}\,\rm{dE} $}}

The normalisation factor ${\rm(\frac{dF}{dE})_{th}}$ can thus be estimated from the N$_\gamma$ measured for each run
and the integral of the differential rate multiplied by the total livetime of
the run T$_{\rm{L}}$.

%%%%%%
\section{Results and Conclusions}
%%%%%%

\begin{figure}
\begin{center}
\includegraphics [width=0.4\textwidth]{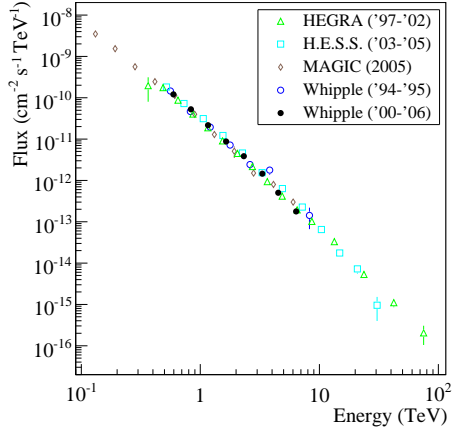}
\vspace{-0.7cm}
\end{center}
\caption{Total 2000--2006 measurement of the Crab Nebula energy spectrum compared to previous measurements.}\label{fig4} 
\end{figure}
%%%%%%

\begin{table}[t]
\begin{center}
\begin{tabular}{cccc}
\hline
\hline
Dataset& $\sigma$/$\sqrt{hr}$ & $(\frac{dF}{dE})_{1~TeV}$  & $\Gamma$ \\
 (Year) &      &  &  \\
\hline                    
'00 - '01 & 5.32 & 3.38 $\pm$ 0.19 & 2.57 $\pm$ 0.10 \\
'01 - '02 & 6.19 & 3.10 $\pm$ 0.11 & 2.54 $\pm$ 0.06 \\
'02 - '03 & 6.91 & 2.75 $\pm$ 0.16 & 2.66 $\pm$ 0.11 \\
'03 - '04 & 5.21 & 3.75 $\pm$ 0.29 & 2.45 $\pm$ 0.10 \\
'04 - '05 & 5.54 & 2.90 $\pm$ 0.26 & 2.68 $\pm$ 0.19 \\
'05 - '06 & 5.28 & 3.58 $\pm$ 0.25 & 2.60 $\pm$ 0.12 \\
\hline
Tot.$_{\rm{Hard}}$  & 5.64 & 3.19 $\pm$ 0.07 & 2.64 $\pm$ 0.03\\
Tot.$_{\rm{Loose}}$  & 3.48 & 3.31 $\pm$ 0.10 & 2.62 $\pm$ 0.04\\
\hline
\end{tabular}
\caption[Energy spectrum of the Crab Nebula measured with the Whipple 10 m]
{Energy spectrum of the Crab Nebula with Hard cuts. The power law flux 
normalization $(\frac{dF}{dE})$ at 1 TeV is in units of \mbox{10$^{-11}$ cm$^{-2}$ s$^{-1}$ TeV$^{-1}$}. 
}\label{TabCrabSpec}
\end{center}
\end{table}

A total Crab Nebula energy spectrum over 2000--2006 was measured 
with both Hard and Loose ~selection cuts. Figure \ref{fig3} shows a contour plot from the $\chi^{2}$ fit errors in flux normalization 
factor ${\rm(\frac{dF}{dE})_{th}}$ and photon index $\Gamma$ for the total Whipple 10 m 2000--2006 dataset with Hard and 
Loose ~cuts. The best fit values agree at the 2$\sigma$ level. Using Hard cuts, the best fit model was a power law over the energy range 
0.49--8 TeV with: {\small \[\frac{\rm{dN}}{\rm{dE}} = (3.19 \pm 0.07)\times10^{-11}\cdot\left(\frac{\rm{E}}{1 
\rm{TeV}}\right)^{-2.64 \pm 0.03}\] }
{\tiny \begin{flushright} $\rm{cm}^{-2} \rm{s}^{-1} \rm{TeV}^{-1}$ \end{flushright}}

%%%%%

Figure \ref{fig4}
shows the corresponding total energy spectrum of the Crab Nebula compared to past measurements with the Whipple 10 m in 
1994--1995, HEGRA in 1997--2002, and H.E.S.S.~in 2003--2005 $[10][2][4]$. 

Table \ref{TabCrabSpec} lists the results from a power law fit 
to the energy spectrum for the individual observing seasons using the Hard cuts (a more
comprehensive table including Loose cuts results can be found in $[5]$). A reasonable agreement in the measured photon index $\Gamma$ and flux 
normalization factor ${\rm(\frac{dF}{dE})_{th}}$ is found between data sets within statistical errors. The RMS spread in 
$\Gamma$ between datasets is 0.06, with a mean statistical error of 0.11 from
each dataset.

\begin{figure}
\begin{center}
\includegraphics [width=0.49\textwidth]{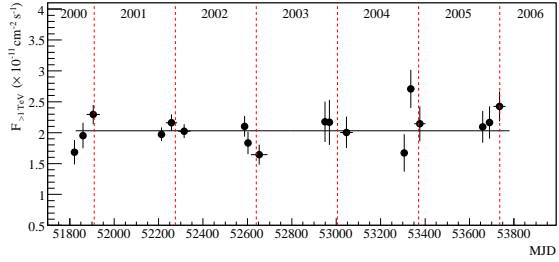}
\vspace{-1.2cm}
\end{center}
\caption{Integral flux of the Crab Nebula over time.}\label{fig5}
\end{figure}

%%%%%%%%%%%%%%%%%%
The integral 
flux  F$_{> 1 \rm{TeV}}$ was calculated from the fitted power law spectrum for
each dataset. As can be seen in figure \ref{fig5}, the measured values were
consistent with steady emission over 6 years; between datasets 
the RMS spread in the integral flux was 12\%. This gives us a sound basis for
the characterisation of the flux variability of other sources. 

%%%%%%

\section{Acknowledgements}{\small 
This research is supported by grants from the U.S. Department of Energy, the
U.S. National Science Foundation, the Smithsonian Institution, by NSERC in
Canada, by PPARC in the UK, and by Science Foundation Ireland.
}
%%%%%% 
\nocite{ref1}
\nocite{ref2}
\nocite{ref3}
\nocite{ref4}
\nocite{ref5}
\nocite{ref6}
\nocite{ref7}
\nocite{ref8}
\nocite{ref9}
\nocite{ref10}
%%%%%%
%This is the reference to .bib file (Whitout .bib!)
{\small
\bibliography{bib}
%This in the bibtex style, is ok.
\bibliographystyle{plain}
}
\end{document}